\documentclass[12pt, letterpaper]{jfm}
\usepackage[utf8]{inputenc}
\usepackage{graphicx} 
\usepackage{epstopdf, epsfig}
\usepackage{gensymb}
\usepackage{siunitx}
\usepackage{upgreek}

\usepackage{amssymb,amsmath}
\usepackage{mathrsfs} 

\graphicspath{ {images/} } 

\shorttitle{Bistability in Polymer Solution Flow Through Porous Media}
\shortauthor{C. A. Browne, A. Shih, S. S. Datta}

\title{Bistability in the Unstable Flow of Polymer Solutions Through Porous Media}
\date{4, February 2020} 

\author{Christopher A. Browne\aff{1},
  Audrey Shih\aff{1},
 \and Sujit S. Datta\aff{1}
 \corresp{\email{ssdatta@princeton.edu}}}

\affiliation{\aff{1}Department of Chemical and Biological Engineering, Princeton University,\\
Princeton, NJ 08544, USA}

\begin{document}
\maketitle

\begin{abstract}
Polymer solutions are often injected in porous media for applications such as oil recovery and groundwater remediation. As the fluid navigates the tortuous pore space, elastic stresses build up, causing the flow to become unstable at sufficiently large injection rates. However, it is poorly understood how the spatial and temporal characteristics of this unstable flow depend on pore space geometry. We elucidate this dependence by systematically varying the spacing between pore constrictions in a model porous medium. We find that when the pore spacing is large, unstable eddies form upstream of each pore, similar to observations of an isolated pore. By contrast, when the pore spacing is sufficiently small, the flow exhibits a surprising bistability, stochastically switching between two distinct unstable flow states. We hypothesize that this unusual behavior arises from the interplay between flow-induced polymer elongation and relaxation of polymers as they are advected through the porous medium. Consistent with this idea, we find that the flow state in a given pore persists for long times. Moreover, we find that the flow state is correlated between neighboring pores; however, these correlations do not persist long-range. Our results thus help to elucidate the rich array of flow behaviors that can arise in polymer solution flow through porous media.	
	
\end{abstract}


\noindent\textit{This preprint was initially submitted on August 6, 2019. An updated version is currently in press at the \textit{Journal of Fluid Mechanics} and will be posted on arxiv in six months.}

\section{Introduction}
\noindent Concentrated polymer solutions have elastic properties that can dramatically alter their flow behavior. Such effects are often harnessed to improve oil recovery and groundwater remediation efforts \citep{sorbie2013, roote1998}. In these cases, an injected polymer solution flows through a tortuous porous medium, such as a reservoir rock or a subsurface aquifer, and displaces trapped non-aqueous fluids from the pore space, enabling them to be recovered downstream. It is therefore critical to be able to predict how the flow behavior depends on the solution properties, injection conditions, and porous medium geometry. Laboratory and field tests provide key empirical measurements of macroscopic variables including fluid pressure and total recovery \citep{sandiford1964, durst1981, pitts1995, wang2011, wei2014, vermolen2014}. However, broad application is still limited by an incomplete understanding of the pore-scale features of the fluid flow. As a result, the mechanisms underlying polymer solution-enhanced fluid recovery are still widely debated \citep{haward2003, odell2006, huh2008, zaitoun1988, zaitoun1998, clarke2016}, and general principles for predicting and controlling the flow are lacking. 

Studies of flow in model microfluidic devices provide valuable insights into the pore-scale flow field. Extensive work has focused on the simplified cases of flow into a single pore constriction or flow into narrow constrictions around a single pillar. In these geometries, the polymer chains are aligned and elongated by the flow, generating upstream recirculating eddies that minimize the extensional stress associated with chain misalignment \citep{batchelor1971, boger1987, mongruel1995, mongruel2003, rodd2007}. Polymer elongation along the curved fluid streamlines also produces normal elastic stresses, which persist for a duration $\lambda$ before relaxing. Thus, at sufficiently large flow speeds, these stresses build up in the solution, further perturbing the flow and causing an elastic instability to arise: an unsteady flow state develops in which the eddies have a broad spectrum of spatial and temporal fluctuations \citep{koelling1991,lanzaro2011,kenney2013, ribeiro2014, lanzaro2015, gulati2015, shi2015, shi2016, lanzaro2017, haward2018, qin2019}. A similar instability also arises in other flows with highly-curved fluid streamlines \citep{pearson1976, larson1992,pakdel1996, mckinley1996, groisman2000, pan2013, galindo2014, sousa2015, haward2016elastic, qin2017}. However, it is unclear how these effects manifest in a porous medium composed of many successive pore constrictions and expansions, known as throats and bodies respectively.

Imaging of flow through one-dimensional (1D) arrays of \textit{widely-spaced} pores consistently demonstrates the formation of unstable eddies upstream of each pore, similar to the case of an isolated constriction \citep{galindo2012,khomami1997,arora2002,kenney2013,shi2015,varshney2017}. By contrast, when the spacing between pores is small, chain elongation may persist across multiple pores as the polymers are advected through the pore space. Thus, memory of strain in one pore may influence the flow in a pore further downstream, potentially providing new spatio-temporal structure to the flow. However, this possibility remains to be explored. Studies conducted at a Reynolds number $\textrm{Re}\sim20$ and an Elasticity number $\textrm{El}\sim1$, and thus also subject to inertial effects, show that decreasing the spacing between pores produces stronger flow fluctuations \citep{shi2016}---providing a clue that polymer memory may indeed influence the flow. Nevertheless, whether and how polymer memory impacts flow through a porous medium has not been fully resolved for the case of $\textrm{Re}\ll1$ and $\textrm{El}\gg1$, in which elastic effects dominate and inertial effects do not also arise. This flow regime is particularly relevant to key applications including oil recovery and groundwater remediation, which can have $\textrm{Re}$ ranging from $\sim10^{-11}$ to $10^{-3}$ and $\textrm{El}$ ranging from $\sim10^{2}$ to $10^{11}$.

Here, we use confocal microscopy to investigate the unstable flow of an elastic polymer solution through model porous media at $\textrm{Re}\ll1$ and $\textrm{El}\gg1$. The media are made of 1D arrays of pore throats, enabling us to directly test the relative importance of polymer memory on the flow by varying the spacing between pore throats. When the pore spacing is large, unstable eddies form upstream of each pore throat, similar to observations of an isolated pore. By contrast, when the pore spacing is sufficiently small, the flow exhibits a surprising bistability. In each pore, the flow persists over long durations in one of two distinct flow states: an eddy-dominated state in which a pair of large unstable eddies forms in the corners of the pore body, and an eddy-free state in which strongly-fluctuating fluid pathlines fill the entire pore body and eddies do not form. We hypothesize that this unusual behavior arises from the interplay between flow-induced polymer elongation, which promotes eddy formation, and relaxation of polymers as they are advected between pores, which enables the eddy-free state to form. Consistent with this idea, we find that the flow state in a given pore persists for long times. In addition, we find that the instantaneous flow state is correlated between neighboring pores; however, these correlations do not persist long-range. Our results thus help to elucidate the rich array of behaviors that can arise for polymer solution flow through porous media.

	\begin{figure}
	\centering
	\includegraphics[angle=0,origin=c, width=\textwidth]{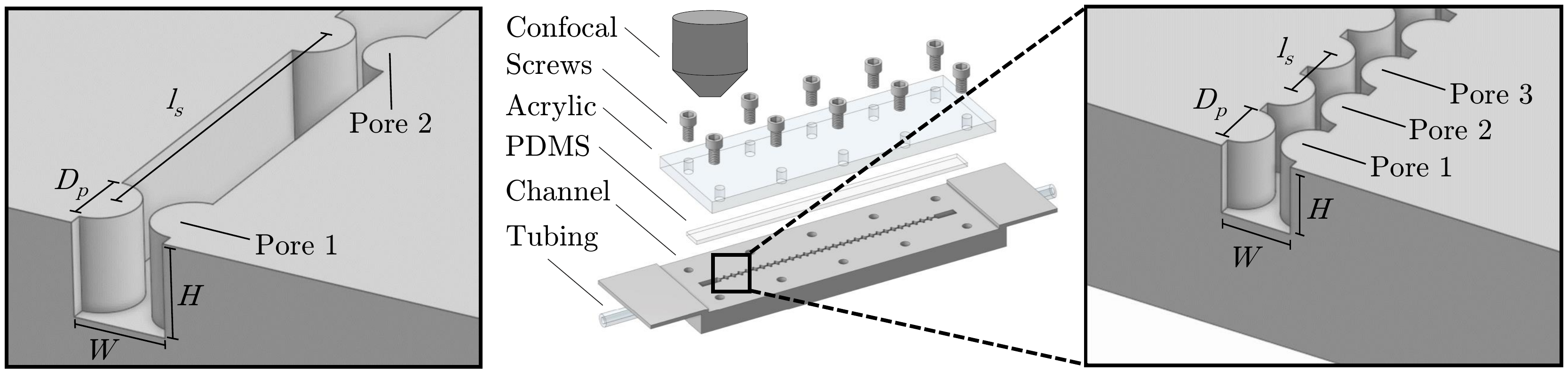}
	\caption{Experimental setup. Fluidic channel contains pore throat constrictions defined by opposing hemi-cylindrical posts and is fabricated using a stereolithographic 3D-printer. Dimensions are $W=2~\textrm{mm}$, $H=2~\textrm{mm}$, $D_\text{p}=1.6~\textrm{mm}$. We vary the pore throat separation distance $l_\text{s}$ and the number of throats in the channel (two examples are shown in the left and right panels). The channel is screwed shut with an acrylic plate over a thin strip of PDMS. Inlet and outlet tubing is glued into 3D-printed holes. The setup is inverted and videos are captured on a confocal microscope; middle panel is vertically flipped for clarity.}
	\label{fig:setup}
\end{figure}

\section{Materials and Methods}
\noindent The void space of a porous medium is typically composed of successive expansions known as pore bodies connected by narrower constrictions known as pore throats \citep{Doyen1988,Kwiecien1990,Bernabe1991,Ioannidis1993}. We use 3D printing to make model porous media that recapitulate these geometric features \citep{leapfrog}. Importantly, this approach provides precise control over the pore space geometry, and yields devices that can be optically interrogated while also withstanding the high pressures that arise during elastic polymer solution flow.

The media are made of straight, square channels having constrictions defined by evenly-spaced hemi-cylindrical posts, as illustrated in Figure \ref{fig:setup}. Each channel is $W=$ $2$ mm wide, $H=$ $2$ mm high, and 7 cm long, with opposing posts of diameter $D_\text{p}=$ $1.6$ mm that are laterally separated by $0.4$ mm and spaced by a center-to-center distance of $l_\text{s}$ along the flow direction. The space between hemi-cylinders along the flow direction thus defines the pore bodies, while the lateral constriction between opposing hemi-cylinders defines the pore throats. Varying $l_\text{s}$ provides a way to systematically test the influence of pore spacing on the flow; hence, our experiments probe an isolated pore throat with $l_\text{s}\rightarrow\infty$, a pair of throats with $l_\text{s}=16W$, and an array of thirty throats with $l_\text{s}=1W$. 

To fabricate each device, we 3D-print the open-faced channel with a FormLabs Form 2 stereolithography printer, using a proprietary clear polymeric resin (FLGPCL04) composed of methacrylate oligomers and photoinitiators. We then glue inlet and outlet tubing directly into 3D-printed connectors designed to minimize perturbation of the polymers away from the pores. Finally, as shown in figure \ref{fig:setup}, the whole assembly is screwed shut using a clear acrylic sheet laser-cut to size and placed on top of a thin strip of polydimethylsiloxane (PDMS), which provides a water-tight seal. 

The fluid used is an aqueous solution of 18 MDa partially hydrolyzed polyacrylamide (HPAM, 30\% carboxylated monomers; Polysciences), a polymer commonly used in oil recovery \citep{sandiford1964, wei2014}. We dissolve 300 ppm HPAM, which corresponds to $\approx 0.3$ the overlap concentration, in a solvent of 10 vol\% ultrapure water (Millipore) and 90 vol\% glycerol (Sigma Aldrich) containing 1 wt\% NaCl (Sigma Aldrich). We also seed the polymer solution with 1 ppm of fluorescent \SI{1}{\micro\meter}  polystyrene tracer particles (Invitrogen) to enable flow visualization. For each porous medium geometry tested, we prepare a fresh solution and use it within one month. 

To visualize the flow, we invert and mount each assembled device on a Nikon A1R inverted laser-scanning confocal microscope. The polymer solution is subsequently injected through the porous medium at a fixed volumetric flow rate $Q$ using a syringe pump (Harvard Apparatus PHD 2000). The time required to inject a single pore throat volume is then given by $\tau_\text{pv}\equiv V_\text{pv}/Q$, where we define the pore throat volume as the void volume between the beginning and end of a pair of opposing hemi-cylinders, $V_\text{pv}= D_\text{p}WH-\pi D_\text{p}^2H/4$. In our experiments at sufficiently large $Q$, we find that unstable fluctuations begin after $\approx10\tau_\text{pv}$ and continue to develop over a time scale of $\approx 40\tau_\text{pv}$, after which it reaches a dynamic equilibrium in which the statistical properties of the flow, described in Section 3, do not appreciably change. Our flow visualization measurements are thus taken $\approx 300\tau_\text{pv}$ after initiating the flow to ensure that the unstable state is fully developed. We acquire fluorescence images every 33 ms from an optical slice of \SI{17.9}{\micro\meter} thickness in the center of the channel height. To visualize the pathlines of individual tracer particles, we average successive frames for 10$\tau_\text{pv}$.

\begin{figure}
	\centering
	\includegraphics[angle=0,origin=c, width=\textwidth]{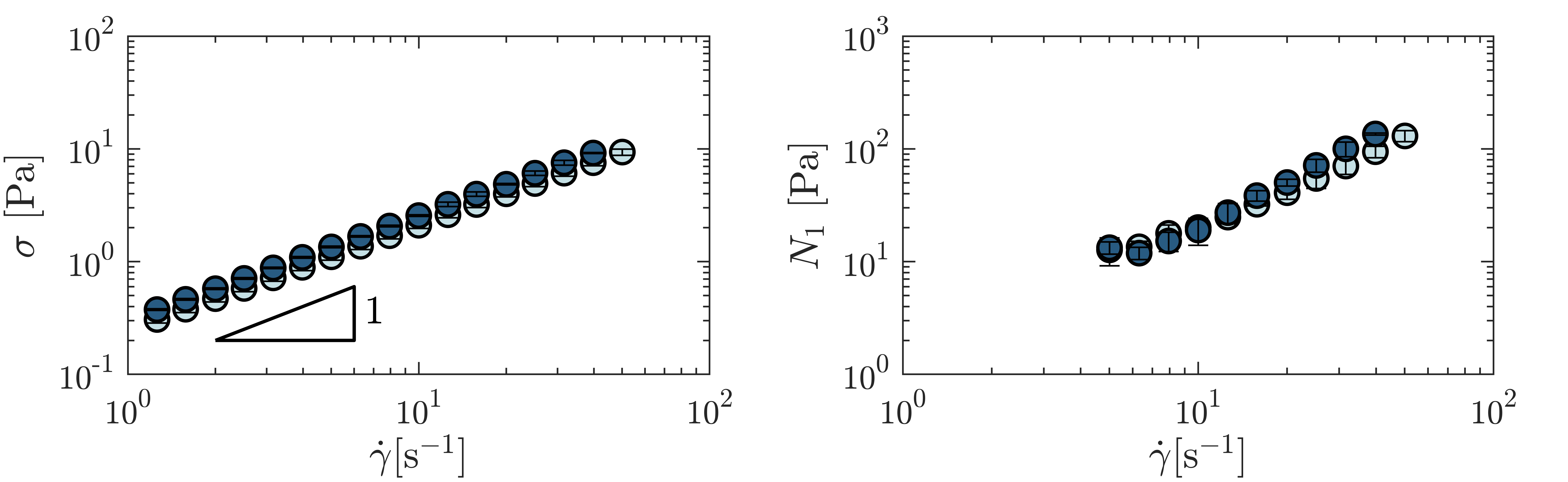}
	\caption{Rheology of HPAM polymer solutions. Left panel shows the shear stress $\sigma$ and right panel shows the first normal stress difference $N_1$ as functions of the shear rate $\dot{\gamma}$ for the sample used for figures \ref{fig:heatmaps}c, \ref{fig:fullPorousStreak}, \ref{fig:temporal}, and \ref{fig:spatial}b--d. Data are taken before (dark blue) and after (light blue) a porous media flow experiment, showing that polymers are not appreciably degraded by the unstable flow. 	Error bars represent standard deviation over five replicate rheological measurements. Fresh polymer solutions are prepared for each experiment, and all dimensionless quantities described in the text are calculated using rheology on fresh samples. }
	\label{fig:rheo}
\end{figure}

We characterize the solution rheology using an Anton Paar MCR-501 rheometer with a 50 mm $1^{\circ}$ cone-plate geometry. Figure \ref{fig:rheo} shows the measured shear stress $\sigma$ and first normal stress difference $N_{1}$ over different shear rates $\dot{\gamma}$ for a representative solution. Furthermore, to assess possible degradation of polymers due to unstable flow in the porous media \citep{vanapalli2006}, we also measure the solution rheology after performing flow experiments at the highest flow rates tested. We find no significant difference between the solution rheology measured before (dark blue points in figure \ref{fig:rheo}) and after (light blue points) flow experiments, indicating minimal polymer degradation due to the unstable flow. For each solution used in each experiment, we obtain at least five replicate measurements of the fluid rheology and use the averaged values in all subsequent calculations.

The shear rate varies approximately linearly with shear rate, with a shear thinning exponent $n\approx0.92$, indicating that shear thinning effects are small due to the high viscosity of the background solvent. Indeed, the pure solvent viscosity is approximately 0.8 times the measured solution viscosity. However, for accuracy, we use the rate-dependent shear viscosity $\mu(\dot{\gamma})\equiv\sigma(\dot{\gamma})/\dot{\gamma}$ in all calculations. To determine the  shear rate that characterizes flow in porous media, we calculate the wall shear rate in the pore throat at each value of $Q$ tested. We do this using the flow profile for a non-Newtonian fluid having the measured shear-thinning exponent $n=0.92$, as calculated by \cite{harnett1979} and by \cite{son2007}. 

We define the Reynolds number comparing inertial to viscous stresses as $\mathrm{Re}\equiv\rho U_\text{t} L/\mu(\dot{\gamma})$, where $\rho$ is the density of the solvent, $U_\text{t}\equiv Q/A_\text{t}$ is the speed corresponding to flow through the pore throat cross-section $A_\text{t}=(W-D_\text{p})H$, and the length scale $L$ is chosen to be half the constriction width $\frac{1}{2}(W-D_\text{p})$. This estimate represents an upper bound for the Reynolds number characterizing the flow; in our porous media experiments, $\mathrm{Re}$ ranges from $\approx8\times10^{-5}$ to $7\times10^{-3}$, indicating that viscous stresses dominate over inertial stresses.

Another common descriptor of elastic flows is the Weissenberg number, which compares elastic stresses to viscous stresses. We define this parameter as $\mathrm{Wi}\equiv N_1(\dot{\gamma})/2\sigma(\dot{\gamma})$, following convention. In our porous media experiments, $\mathrm{Wi}$ ranges from $\approx2$ to 9, indicating that elastic stresses dominate. Moreover, the corresponding values of the Elasticity number $\mathrm{El}\equiv\mathrm{Wi}/\mathrm{Re}$, which compares elastic stresses to inertial stresses, are $\gtrsim900$. Our experiments thus probe the elasticity-dominated flow regime. 

An elastic instability arises when elastic stresses---characterized by a large value of $\mathrm{Wi}$---persist over a polymer relaxation length scale $\lambda U_\text{t}$ exceeding an effective streamline radius of curvature $\mathscr{R}$. Thus, unstable flow is predicted to arise for sufficiently large values of the parameter $\mathrm{M}\equiv\sqrt{2\mathrm{Wi}~\lambda U_\text{t}/\mathscr{R}}$, as confirmed experimentally for diverse flow geometries \citep{pakdel1996, mckinley1996,haward2016elastic,zilz2012}. Indeed, M quantifies the largest destabilizing term in the Navier-Stokes equations for elastic flows \citep{pakdel1996, mckinley1996}; we thus use this parameter to describe the different flow regimes tested in our experiments. To compute this parameter, we use the rheology measurements to calculate the relaxation time $\lambda=\mathrm{Wi}/\dot{\gamma}$, and use an empirical fit previously established by \cite{mckinley1996} to calculate the radius of curvature $\mathscr{R}\approx\left(2/D_\text{p}+32.5/W\right)^{-1}$. We find $\lambda\approx0.3$ to 6 s, in good agreement with previous measurements performed on similar solutions \citep{qin2017}. The corresponding values of M range from $\approx6$ to 31 for our experiments studying flow in porous media.

\section{Results}
\subsection{Isolated pore throat}
\begin{figure}
	\centering
	\includegraphics[angle=0,origin=c, width=\textwidth]{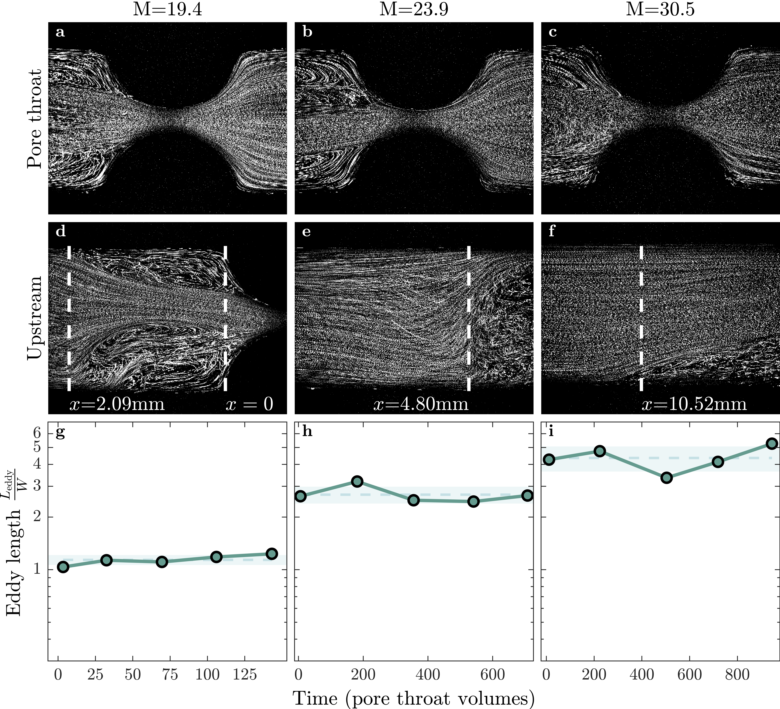}
	\caption{Pathline images of polymer solution flow through a straight channel with a single pore throat. Pathline images are averaged over $10\tau_\text{pv}$. Flow rates are expressed via the $\mathrm{M}$ parameter, which equals 19.4, 23.9, and 30.5 for panels \textbf{a}, \textbf{b}, and \textbf{c} respectively, corresponding to Weissenberg numbers $\mathrm{Wi}=5.6$, 6.9, and 8.8. \textbf{a--c} show strong recirculating eddies upstream of the pore throat, but no downstream eddies at any flow rate. \textbf{d--f} show the leftmost edge of the eddies in front of the pore throat. Dashed lines indicate where the eddy-dominated region begins. $x$ indicates the distance in millimeters from the base of the hemi-cylinder. \textbf{g--i} show the measured eddy length over time normalized by $\tau_\text{pv}$. Shaded regions show standard deviation around the temporal mean (dashed lines).}
	\label{fig:singlestreak}
\end{figure}

\noindent To test the limiting case of widely-spaced pores ($l_\text{s}\rightarrow\infty$), we first investigate flow through an isolated pore throat centered in a channel. At low imposed flow rates, and thus at low values of M, the flow is laminar: the fluid pathlines do not cross and do not change in time. We do not observe eddies---instead, the pathlines smoothly converge as they approach the throat and symmetrically diverge as they leave it. However, above a threshold value of $\mathrm{M}\approx19$, we observe the onset of a flow instability: a pair of unstable eddies forms against the channel walls upstream of the pore throat, as exemplified in figures \ref{fig:singlestreak}a-c (at the pore throat) and d-f (upstream) for $\mathrm{M}=19.4$, 23.9, and 30.5. 

Within each eddy, the fluid recirculates with a speed slower than the mean imposed flow speed in the channel. The fluid pathlines continually fluctuate on long timescales, and also continually cross, indicating that fluctuations in the flow occur on a time scale shorter than the pathline duration of 10$\tau_\text{pv}$ (figures \ref{fig:singlestreak}d-f). These fluctuations are reflected in the motions of the eddy boundaries and lengths, which also fluctuate as the flow progresses. Similarly, while the fluid in the region formed between the eddies does not recirculate, the fluid pathlines also continually fluctuate on long timescales and continually cross, reflecting the presence of rapid fluctuations in the flow throughout. By contrast, the flow is more steady downstream of the pore throat: we do not observe any eddies or marked temporal changes in the flow for all values of M tested. This result is consistent with the findings of \cite{qin2019}, who also found highly unstable eddies upstream of a cylinder in a channel, but suppressed fluctuations and no eddies downstream. We therefore focus our subsequent analysis on the upstream region.

To further characterize this behavior, we track the eddy length $L_\text{eddy}$ over time for each value of M tested. We measure $L_\text{eddy}$ from the base of the hemi-cylinders ($x=0$ in figure \ref{fig:singlestreak}d) to the farthest upstream location having a pathline oriented perpendicular to the imposed flow direction (dashed lines in figures \ref{fig:singlestreak}d-f). Consistent with the visual observations, $L_\text{eddy}$ fluctuates over time in each experiment, as indicated by the shaded regions in figures \ref{fig:singlestreak}g-i; however, it fluctuates around a single mean value that increases with M. We quantify these fluctuations using the coefficient of variation $c_{\textrm{v}}$, defined as the ratio between the standard deviation and the mean of the measurements of $L_\text{eddy}$ over time. Taking data from two replicate experiments at these imposed flow rates, we find $c_{\textrm{v}}\approx0.3$, 0.3, and 0.4 for $\mathrm{M}=19.4$, 23.9, and 30.5 respectively. 

\begin{figure}
	\centering
	\includegraphics[angle=0,origin=c, width=\textwidth]{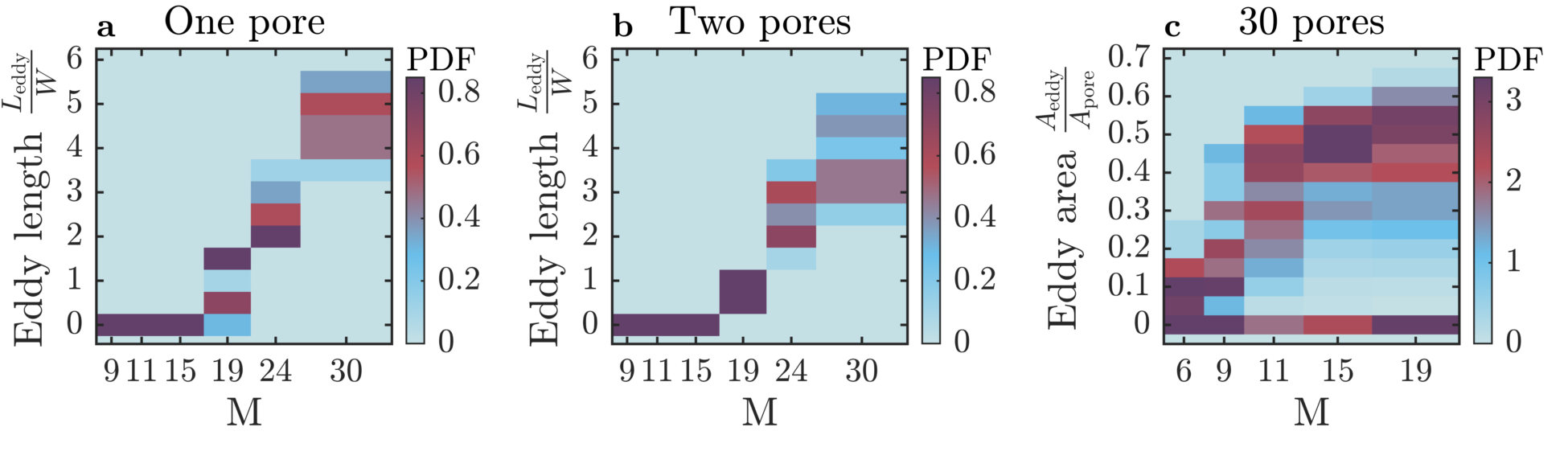}
	\caption{Probability density functions of measured eddy sizes for different values of M, averaged over at least $7\tau_\text{pv}$ and over two separate replicate experiments, displayed vertically as heatmaps. \textbf{a} PDFs of $L_\text{eddy}/W$ for a channel with a single pore throat constriction. For each value of M there is a single characteristic eddy length that increases monotonically with M. \textbf{b} PDFs of $L_\text{eddy}/W$ for a channel with two widely-spaced pores $l_\text{s}=16W$ apart. Again, for each value of M there is a single characteristic eddy length that increases monotonically with M. \textbf{c} PDFs of $A_\text{total}/A_\text{pore}$ for a porous medium with 30 closely-spaced pores $l_\text{s}=W$ apart. Here, the PDFs for $\textrm{M}>9$ are bimodal, showing multiple characteristic eddy areas (one peak at $\frac{A_{\textrm{eddy}}}{A_{\textrm{pore}}}\approx60\%$ and one peak at $\frac{A_{\textrm{eddy}}}{A_{\textrm{pore}}}\approx0$). The two branches in the PDFs indicate a bistability in unstable flow states.}
	\label{fig:heatmaps}
\end{figure}

We summarize all of our measurements by plotting the probability density (PDF) of time-averaged measured eddy lengths for each value of M tested. Below the threshold value of $\mathrm{M}\approx19$, we do not observe eddies and hence $L_\text{eddy}=0$. By contrast, above this threshold, $L_\text{eddy}>0$ fluctuates about a well-defined mean value, which increases with M, as shown in figure \ref{fig:heatmaps}a. This increase in $L_\text{eddy}$ is similar to previous measurements for isolated constrictions; these studies demonstrate that eddies form when polymers are elongated, and the size of eddies grows as polymers are increasingly elongated \citep{batchelor1971, boger1987, mongruel1995, mongruel2003, rodd2007}. Our results thus suggest that flow fluctuations arising from unstable flow elongate the individual polymer chains \citep{sureshkumar1997, balkovsky2000, chertkov2000, gupta2004, terrapon2004}, which then generate unstable upstream eddies to minimize extensional stresses \citep{batchelor1971, boger1987,mongruel1995, mongruel2003,rodd2007}.

\subsection{Two widely-spaced pore throats}
\begin{figure}
	\centering
	\includegraphics[angle=0,origin=c, width=\textwidth]{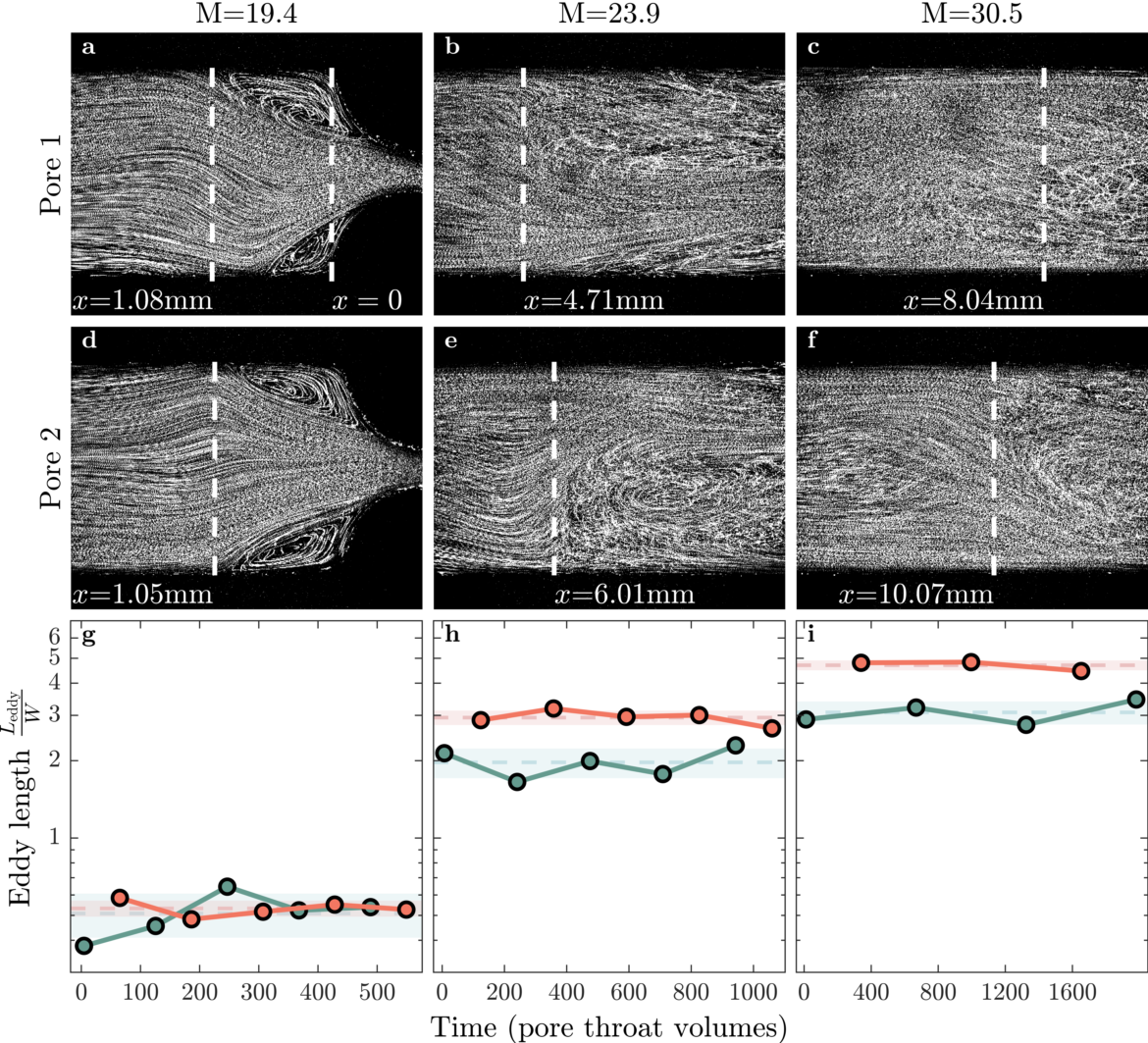}
	\caption{Pathline images of polymer solution flow through a straight channel with two pore throats separated by $l_\text{s}=16W$. Pathline images are averaged over $10\tau_\text{pv}$. Flow rates are expressed via the $\mathrm{M}$ parameter, which equals 19.4, 23.9, and 30.5 for \textbf{a}, \textbf{b}, and \textbf{c} respectively, corresponding to $\mathrm{Wi}=5.6$, 6.9, and 8.8. \textbf{a--c} show the leftmost edge of the eddies in front of the first pore throat. \textbf{d--f} show the leftmost edge of the eddies in front of the second pore throat. No eddies are observed downstream of either throat. \textbf{g--i} show the measured eddy lengths over time normalized by $\tau_\text{pv}$ for pore 1 (green) and pore 2 (red). Shaded regions show standard deviation around the temporal mean (dashed lines).	}
	\label{fig:doublestreak}
\end{figure}

\noindent We next investigate two pore throats spaced a distance $l_s=16W$ apart along the flow direction. As we find with an isolated throat, the flow is laminar at low values of M, while above a similar threshold value of $\textrm{M}\approx19$, we observe the onset of the flow instability. A pair of unstable eddies again forms against the channel walls upstream of each pore throat, as exemplified in figures \ref{fig:doublestreak}a-c (first  throat) and d-f (second throat). Moreover, as with the isolated pore throat, the flow is more steady immediately downstream of each throat, with no observable eddies or temporal changes in the flow for any values of M tested.

We again quantify this behavior by measuring $L_\text{eddy}$ over time for each value of M tested. For each pore throat, $L_\text{eddy}$ again fluctuates around a single mean value that increases with M (figures \ref{fig:doublestreak}g-i)---similar to the case of an isolated throat. The PDFs of the combined time-averaged measurements of $L_\text{eddy}$ also reflect its increase with M, as shown in figure \ref{fig:heatmaps}b. Intriguingly, however, we observe two key differences from the isolated throat. First, while the mean values of $L_\text{eddy}$ are similar for the two pore throats, eddies upstream of the second throat (red points, figures \ref{fig:doublestreak}g-i) are slightly larger than eddies upstream of the first throat (green points) for large values of M. Second, the eddies upstream of the second throat are less unstable---the temporal fluctuations in $L_\text{eddy}$ are notably suppressed for the second pore throat for all M above the threshold for unstable flow (compare red to green shaded regions in figures \ref{fig:doublestreak}g-i). Comparing the coefficients of variation confirms this finding: for the first pair of eddies, $c_{\textrm{v}}=0.2\pm0.1$ while for the second pair, $c_{\textrm{v}}=0.10\pm0.02$, which is significantly smaller ($p=0.02$, one-tailed $t$-test). Thus, when the spacing between pore throats is reduced, the spatio-temporal characteristics of the flow are altered---presumably because polymer elongation can persist across multiple pores.

\begin{figure}
	\centering
	\includegraphics[angle=0,origin=c, width= \textwidth]{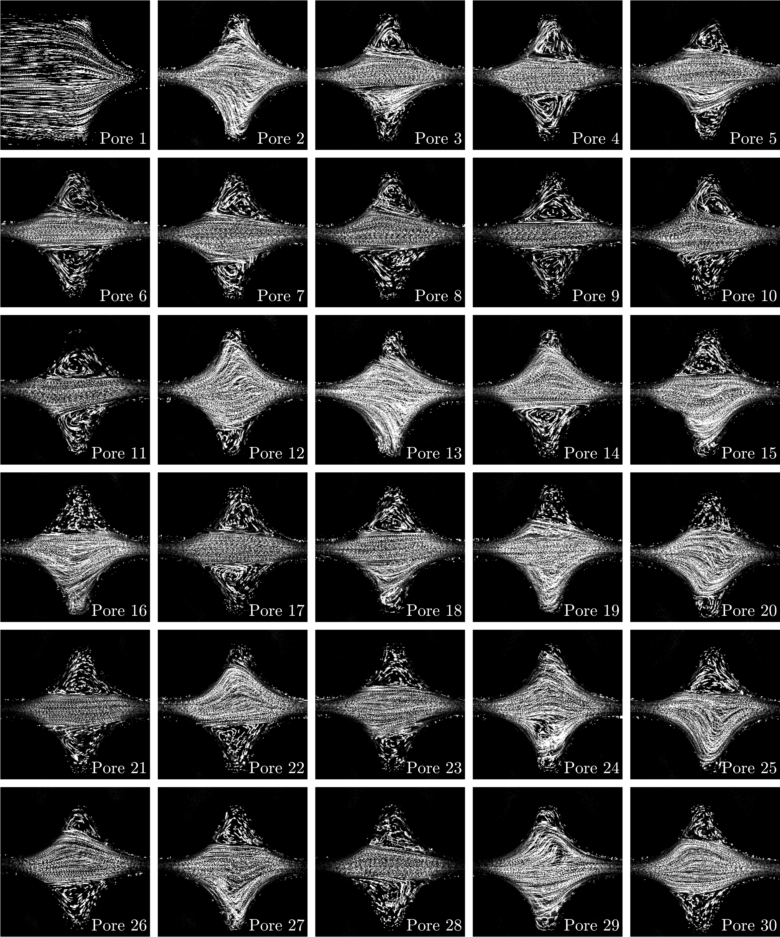}
	\caption{Pathline images of polymer solution flow through a porous medium with 30 pore throats separated by $l_\text{s}=W=2$ mm at $\mathrm{M}=11.4$. Images span 2.11 mm across. Pathline images are averaged over $10\tau_\text{pv}$. Pore 6 exemplifies the eddy-dominated state, with eddies that continually fluctuate both internally and at their boundaries, while pore 2 exemplifies the eddy-free state, with strongly-fluctuating pathlines that fill the entire pore body. }
	\label{fig:fullPorousStreak}
\end{figure}

\subsection{Thirty closely-spaced pore throats}

\noindent To further test the hypothesis that polymer memory impacts flow behavior, we next investigate flow through a medium with an even smaller spacing between pore throats. Specifically, the flow channel contains thirty pore throats spaced a distance $l_\text{s}=W$ apart along the flow direction. In this case, we find that the flow behavior is strikingly different from the larger $l_\text{s}$ cases described in Sections 3.1 and 3.2. 

One key difference is that the threshold for the onset of the flow instability is dramatically lowered. At low imposed flow rates, and thus at low values of M, the flow is laminar: the fluid pathlines do not cross and do not vary over time. In this regime, a pair of small, symmetric, laminar, recirculating eddies forms in the corners of each pore body due to the small spacing between successive pore expansions and constrictions. Above a threshold value of $\mathrm{M}\approx9$---considerably smaller than the threshold $\mathrm{M}\approx19$ for the single- and double-throat cases---we observe the onset of unstable flow: the fluid pathlines continually cross and vary over time. Thus, decreasing the spacing between pore throats decreases the threshold value of M required for unstable flow, suggesting that polymer memory strongly impacts the flow behavior. 

Even more strikingly, we observe two distinct flow states that can arise in each pore body throughout the medium: an `eddy-dominated' state in which a pair of large unstable eddies forms in the corners of the pore body, and an `eddy-free' state in which strongly-fluctuating fluid pathlines fill the entire pore body and eddies do not form. This surprising bistability is illustrated in figure \ref{fig:fullPorousStreak}, which shows the pathline images taken sequentially from each pore in the medium at $\mathrm{M}=11.4$. Pore 6 exemplifies the eddy-dominated state, with eddies that continually fluctuate both internally and at their boundaries, while pore 2 exemplifies the eddy-free state, with strongly-fluctuating pathlines that fill the entire pore body. Though these snapshots are taken at an optical slice in the center of the channel height, imaging at other heights shows similar flow pathlines, indicating that the spatial structure of the flow does not vary appreciably across the channel height. This observation of distinct pore-scale flow states is in stark contrast to the typical assumption that the spatio-temporal characteristics of unstable flow do not vary through a porous medium: typically no differentiation is made between the flow behaviors that manifest in the different pores.

\begin{figure}
	\centering
	\includegraphics[angle=0,origin=c, width= \textwidth]{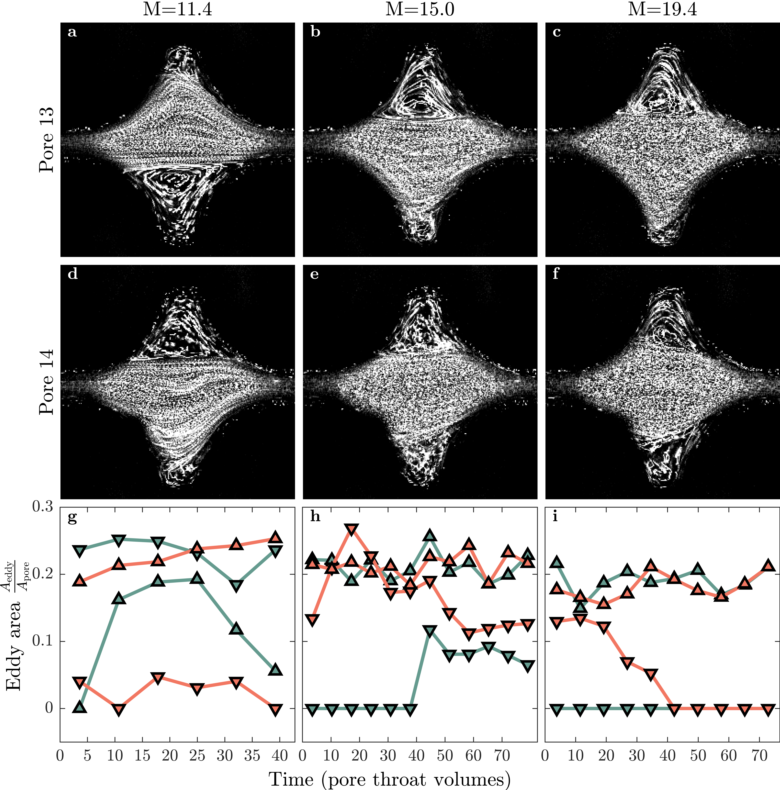}
	\caption{Pathline images of polymer solution flow through pores 13 \textbf{a--c} and 14  \textbf{d--f} at different flow rates corresponding to $\mathrm{M}=11.4$, 15, and  19.4 (left to right). Pathline images are averaged over $10\tau_\text{pv}$. \textbf{g--i} show the measured eddy areas $A_\text{eddy}/A_\text{pore}$ over time normalized by $\tau_\text{pv}$ for pore 13 (green) and pore 14 (red). Upward triangles are for the top region of the pore body, downward triangles are for the bottom region of the pore body. }
	\label{fig:temporal}
\end{figure}

To quantify this behavior, we measure the two-dimensional (2D) area of the individual eddies $A_\text{eddy}$ over time for each value of M tested. In laminar flow, the eddies occupy only $\approx5\%$ of the total area of a pore, which we define as $A_\text{pore}\equiv Wl_\text{s}-\pi D_\text{p}^{2}/4$. By contrast, unstable eddies in the eddy-dominated state have values of $A_\text{eddy}$ that fluctuate strongly in time, and whose mean value can be up to $\approx30\%$ of $A_\text{pore}$, while in the eddy-free state $A_\text{eddy}\approx0$. Intriguingly, while the two flow states are each unstable, the flow behavior in each pore body appears to be bistable, as illustrated in figure \ref{fig:temporal}: a pore will persist in a given unstable flow state for a long duration of time before switching, seemingly randomly, to the other flow state. For example, at $\text{M}=11.4$, the upper eddy in pore 13 and the lower eddy in pore 14 (figure \ref{fig:temporal}g, upward-pointing red triangles and downward-pointing green triangles respectively) persist in the eddy-dominated state over the entire imaging duration, while the lower eddy in pore 13 persists in the eddy-free state (downward red triangles). However, the upper eddy in pore 14 initially switches from the eddy-free to the eddy-dominated state, in which it persists for 15$\tau_\text{pv}$ before switching back to the eddy-free state (upward green triangles). We observe this flow bistability in all pores of the medium, and at all values of M tested; two more examples for M = 15 and 19.4 are shown in figures \ref{fig:temporal}h-i respectively.

To further characterize the flow bistability shown in figures \ref{fig:temporal}g-i, we plot the PDFs of the time-averaged measurements of $A_\text{total}$, which describes the total $A_\text{eddy}$ measured in each pore combined for all thirty pores in the medium. Below $\mathrm{M}\approx9$, $A_\text{total}\approx10\%$ of $A_\text{pore}$, and eddies do not change in time; by contrast, above the onset of the flow instability at $\mathrm{M}\approx9$, the PDFs become bimodal, reflecting the bistability in flow behavior (figure \ref{fig:heatmaps}c). The eddy-dominated state is represented by the upper branch of the PDFs, in which $A_\text{total}$ increases with M, eventually plateauing at $\approx60\%$ of $A_\text{pore}$ at the highest values of M tested. The eddy-free state is represented by the lower branch of the PDFs, in which $A_\text{total}\approx0$ over all M. This bistability does not arise in porous media with wider pore spacings: the lower branch of the PDFs does not appear in figures \ref{fig:heatmaps}a-b. Thus, when the spacing between pore throats is reduced from $l_\text{s}=16W$ to $l_\text{s}=W$---and thus, elongation of individual polymers is more likely to persist across multiple pores---the flow abruptly becomes bistable, exhibiting two coexisting unstable flow states. Indeed, previous work has theorized that unstable flow may bifurcate into two coexisting flow states \citep{avgousti1993, sureshkumar1994, varshney2017}; to our knowledge, our work is the first experimental confirmation of this prediction.

\subsubsection{Bistability and polymer conformations in flow}
\noindent How does this unusual flow bistability arise? Previous measurements of polymer conformations indicate that differing fractions of coiled and elongated chains coexist in extensional and unstable flows depending on the imposed flow conditions \citep{schroeder2003, gerashchenko2005, francois2009}. Furthermore, simulations indicate that polymers having different elongation can have dramatically differing pore-scale flow behaviors \citep{pilitsis1989, pilitsis1991}. Thus, we hypothesize that flow bistability arises from the interplay between flow-induced polymer elongation, which promotes eddy formation, and relaxation of polymers as they are advected between pores, which enables the eddy-free state to form. 

We first consider a pore in the eddy-dominated state. Polymers entering the pore are likely in an elongated conformation due to the combined influence of unstable flow fluctuations and extension by flow converging into the upstream pore throat. Indeed, previous work has demonstrated that eddies form upstream of a constriction when polymers are elongated \citep{batchelor1971, boger1987, mongruel1995, mongruel2003, rodd2007}. Eddy formation minimizes extensional stresses in the center of the pore: the net flow through the pore body occurs in a nearly-straight channel spanning one pore throat to the next, as can be seen in the example of pore 6 in figure \ref{fig:fullPorousStreak}. For simplicity, we consider the limit of large M, in which eddies completely fill the corners of the pore body. With the exception of unstable fluctuations, the flow velocities in this channel are then aligned along the flow direction with speed $\sim U_\text{t}\equiv Q/A_\text{t}\sim1$ mm/s, and therefore the extensional component of the flow in the channel is minimal. The elongated polymers thus continue to relax as they are advected through this channel, reaching their equilibrium coiled conformation after a duration $\sim\lambda_\text{rel}$, the chain relaxation time. We compare this time scale to the residence time required for the polymers to transit across the pore body from the upstream throat to the downstream throat, $\sim l_\text{s}/U_\text{t}$, yielding an advective Deborah number $\mathrm{De_\text{adv}}\equiv\lambda_\text{rel}U_\text{t}/l_\text{s}$. When $l_\text{s}$ is small and $\mathrm{De_\text{adv}}\gtrsim1$, chains are still elongated as they enter the next pore body, thereby promoting eddy formation in the current pore and the downstream pore as well. For our experiments in the unstable regime with $l_\text{s}=W$, $\mathrm{De}_\text{adv}$ ranges from $\sim0.5$ to $3$ using $\lambda_\text{rel}\sim\lambda\approx1$ s; this estimate likely under-estimates $\mathrm{De}_\text{adv}$, since $\lambda_\text{rel}$ is known to increase considerably in extensional flow \citep{clasen2006}. Thus, we expect the eddy-dominated state to persist in the pore body over time before random flow fluctuations cause it to switch to the eddy-free state, consistent with our measurements shown in figures \ref{fig:temporal}g-i. We also expect the eddy-dominated state to be correlated between neighboring pores.

We next consider the formation of the eddy-free state. As polymers pass through eddy-dominated pores, they gradually relax to the coiled conformation. When a sufficient fraction of coiled polymers are at the entrance to a pore, there will be no driving force for eddy formation. The pore will thus be in the eddy-free state. The fluid streamlines then diverge from the upstream pore throat, creating a compressional flow that further promotes the coiled conformation. As the polymers continue to traverse the pore body, they remain coiled until they encounter the converging flow into the downstream pore throat. This extensional flow partially elongates the chains, which requires a time scale $\sim\lambda_\text{ret}$. We again compare this time scale to the residence time required for the polymers to transit from the beginning of the converging region to the downstream throat, in this case $\sim H\left(WD_\text{p}/2-\pi D_\text{p}^2/8\right)/Q$. This comparison yields another advective Deborah number $\text{De}'_\text{adv}\equiv\lambda_\text{ret}Q/ \left[H(WD_\text{p}/2-\pi D_\text{p}^2/8)\right]$. When $\text{De}'_\text{adv}\gtrsim1$, chains are not elongated as they enter the next pore body, thereby promoting the eddy-free state in the current pore and the downstream pore as well. For our experiments in the unstable regime, $\text{De}'_\text{adv}$ ranges from $\approx0.7$ to 5 using $\lambda_\text{ret}\sim\lambda\approx1$ s, although these values again likely under-estimate $\text{De}'_\text{adv}$. Thus, similar to the eddy-dominated state, we expect the eddy-free state to persist in the pore body over time before random flow fluctuations cause it to switch to the eddy-free state, consistent with our measurements shown in figures \ref{fig:temporal}g-i. Moreover, we again expect the eddy-free state to be correlated between neighboring pores.

\begin{figure}
	\centering
	\includegraphics[angle=0,origin=c, width= \textwidth]{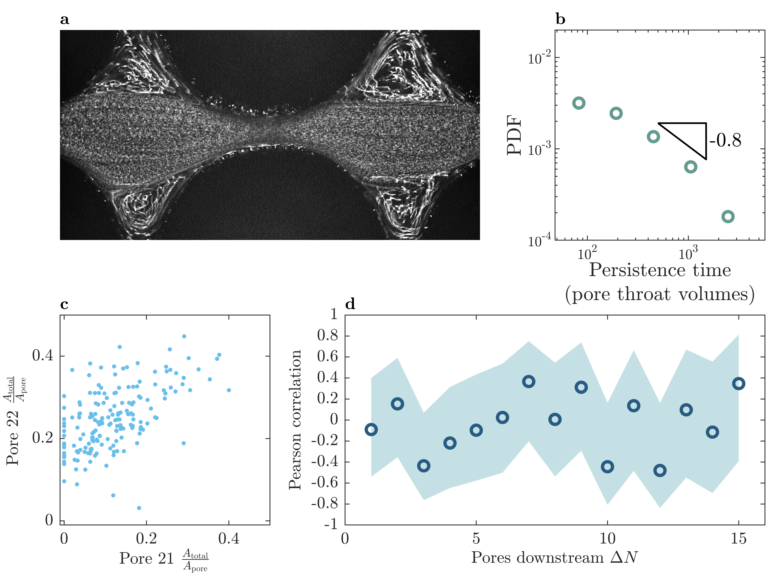}
	\caption{\textbf{a} Simultaneous pathline imaging of pores 21 and 22 averaged over $10\tau_\text{pv}$. Both pores are in the eddy-dominated state. \textbf{b} Probability density function of times between discrete switching events, representing the duration over which a given flow state persists, taken up to $2500\tau_\text{pv}$. The distribution has a long tail, consistent with a power law that decays as $\approx-0.8$. \textbf{c} Instantaneous eddy areas of pore 21 and pore 22 are positively correlated, indicating that flow states are correlated between neighboring pores. \textbf{d} Pearson correlation coefficients between pores separated by a length $l_\text{s}\Updelta N$ and imaged a time $80\tau_\text{pv}\Updelta N$ apart for $\text{M}=15.0$. The flow state of each pore is averaged over $80\tau_\text{pv}$. Light blue shaded region indicates 99\% confidence interval for estimated correlation coefficient given the sampling size.}
	\label{fig:spatial}
\end{figure}

\subsubsection{Temporal and Spatial Characteristics of the Flow}\label{sec:correlation}
\noindent The hypothesis presented in Section 3.3.1 makes two testable predictions: first, that the two different unstable flow states each persist over long times before randomly switching, and second, that the flow states between neighboring pores are correlated. We test these predictions by investigating the temporal and spatial characteristics of the pore-scale flow. Specifically, we simultaneously image two neighboring pores within the medium (pores 21 and 22) and monitor their flow states for $2500\tau_\text{pv}$. Figure \ref{fig:spatial}a shows a snapshot of the flow pathlines imaged within both pores simultaneously at an instance when both pores are in the eddy-dominated state. 

The data in figures \ref{fig:temporal}g--i support the first prediction that the two distinct flow states persist over long times. To further test this prediction, we measure the distribution of  durations over which each flow state persists in a given pore before switching to the other state; we define a switching event as the instant when the total eddy area $A_{\mathrm{total}}$ in a given pore crosses a threshold value of $0.15~A_{\mathrm{pore}}$. This threshold is motivated by the clear separation between the eddy-dominated and eddy-free states indicated by the probability density functions in figure \ref{fig:heatmaps}c. In agreement with our expectation, we find a broad distribution of long flow persistence times, as shown in figure \ref{fig:spatial}b. Intriguingly, the decay of the measured probability density function is consistent with a power law of exponent $\approx-0.8$; thus, unstable flow does not appear to have a characteristic persistence time, and instead can persist for longer than $2500\tau_\text{pv}$, confirming our first prediction.

The image in figure \ref{fig:spatial}a supports the second prediction that the flow states in neighboring pores are correlated. To further test this prediction, we measure the instantaneous total eddy area $A_\text{total}$ in both pores 21 and 22. These measurements, spanning several thousand $\tau_\text{pv}$, are shown in figure \ref{fig:spatial}c. We find that $A_\text{total}$ for pore 22 is positively correlated with $A_\text{total}$ for pore 21; the Pearson correlation coefficient is $\rho_{A_{21},A_{22}}=0.55$, confirming a positive correlation between both quantities that is statistically significant ($p<0.001$, two-tailed $t$-test). Thus, the unstable flow states in neighboring pores are correlated with each other, confirming our second prediction.

However, imaging of flow through the entire medium (figure \ref{fig:fullPorousStreak}) suggests that these spatial and temporal correlations across neighboring pores are not sufficient to produce long-range correlations in the flow over long time scales. To test whether such long-range correlations persist, we sequentially measure the total eddy areas in each pore, each for $80\tau_\text{pv}$, throughout the entire medium. We then calculate the Pearson correlation coefficient between the values of $A_\text{total}/A_\text{pore}$ measured in pores that are separated by a length $l_\text{s}\Updelta N$, where $\Updelta N$ ranges from 1 to 15. We find no long-range correlations in the flow over these long time scales: the Pearson correlation coefficient does not statistically deviate from zero for any values of $\Updelta N$ throughout the medium, as shown in figure \ref{fig:spatial}d. Thus, while unstable flow states are correlated over long times across neighboring pores, these correlations do not persist throughout the entire medium---presumably due to the influence of random fluctuations in the flow. We note, however, that our data do not preclude the possibility that long-range correlations may arise at shorter time scales; our imaging only probes correlations between pores $l_\text{s}\Updelta N$ apart that can persist over a time scale exceeding $\approx80\Updelta N\tau_\text{pv}$. Investigating whether the flow correlations described in figure \ref{fig:spatial}a-c can persist over broader spatial and temporal ranges will be a useful direction for future work.

\section{Conclusions}
\noindent Our work describes the first experimental observations of bistability in the flow of an elastic polymer solution through porous media, confirming previous theoretical predictions \citep{avgousti1993, sureshkumar1994, varshney2017}. We find that when the spacing between pores is sufficiently small, and when the imposed flow rate is sufficiently large, the flow stochastically switches between two distinct unstable flow states. In the eddy-dominated state, a pair of large unstable eddies forms in the corners of a pore body, while in the eddy-free state, strongly-fluctuating fluid pathlines fill the entire pore body and eddies do not form. Our work thus indicates that in a porous medium the pore-scale flow behavior might not be uniquely determined by the M parameter \citep{pakdel1996, mckinley1996} or the Weissenberg number Wi, as is often assumed. Extending these findings to 2D and 3D media, in which transverse interactions between pores may also play a role \citep{talwar1995,khomami1997, arora2002, sadanandan2004, vazquez2012,howe2015, clarke2016, kawale2017a, kawale2017b,de2016,de2017b,de2017a,de2017viscoelastic,de2018flow,de2018viscoelastic,walkama2019}, will be an important direction for future studies.

We hypothesize that the flow bistability arises from the interplay between flow-induced polymer elongation, which promotes eddy formation, and relaxation of polymers as they are advected between pores, which enables the eddy-free state to form. Consistent with this idea, we find that the eddy sizes increase with the imposed flow rate. Additionally, we find that a flow state in a given pore persists over long time scales before switching to the other flow state, presumably due to random flow fluctuations. Flow state is also strongly correlated between neighboring pores; however, these correlations are not sufficient to produce long-range correlations in the flow through the entire medium over long time scales. Elucidating the factors that determine the length and time scales over which flow correlations persist will be an interesting direction for future work.  

Indeed, the different spatial and temporal characteristics of the two flow states could impact fluid mixing and the displacement of trapped immiscible fluids from the pore space in a variety of ways \citep{babayekhorasani2016, aramideh2019}. Thus, a deeper understanding of the flow behaviors discovered here could provide guidance to applications that require specific mixing or fluid displacement behaviors. Examples include oil recovery and groundwater remediation, in which the viscous forces exerted by the polymer solution could displace a trapped fluid from the pores, or unstable mixing due to the fluid instability could improve the transport of oxidants and surfactants to the fluid interface. Such flow behaviors could also be harnessed in other emerging applications such as controlling mixing in lab-on-a-chip devices.\\

It is a pleasure to acknowledge P. D. Olmsted, A. Z. Panagiotopoulos, R. K. Prud'homme, B. Qin, and H. A. Stone for stimulating discussions, and the Stone Lab for access to the SLA printer and the rheometer. Acknowledgment is made to the Donors of the American Chemical Society Petroleum Research Fund for partial support of this research. This material is also based upon work supported by the National Science Foundation Graduate Research Fellowship Program (to C.A.B.) under Grant No. DGE-1656466. Any opinions, findings, and conclusions or recommendations expressed in this material are those of the authors and do not necessarily reflect the views of the National Science Foundation. C.A.B. was also supported in part by the Mary and Randall Hack Graduate Award of the Princeton Environmental Institute. A.S. was supported in part by the Lidow Thesis Fund at Princeton University and the Dede T. Bartlett P03 Fund for Student Research through the Andlinger Center for Energy and the Environment.

\bibliographystyle{jfm}
\bibliography{CB_bibtex_long}

\begin{thebibliography}{81}
\expandafter\ifx\csname natexlab\endcsname\relax\def\natexlab#1{#1}\fi
\def\au#1{#1} \def\ed#1{#1} \def\yr#1{#1}\def\at#1{#1}\def\jt#1{\textit{#1}}
  \def\bt#1{#1}\def\bvol#1{\textbf{#1}} \def\vol#1{#1} \def\pg#1{#1}
  \def\publ#1{#1}\def\arxiv#1{#1}\def\org#1{#1}\def\st#1{\textit{#1}}

\bibitem[Aramideh {\em et~al.\/}(2019)Aramideh, Vlachos \&
  Ardekani]{aramideh2019}
{\sc \au{Aramideh, Soroush}, \au{Vlachos, Pavlos~P} \& \au{Ardekani, Arezoo~M}}
  \yr{2019}  \at{Nanoparticle dispersion in porous media in viscoelastic
  polymer solutions}.  \jt{Journal of Non-Newtonian Fluid Mechanics}
  \bvol{268},  \pg{75--80}.

\bibitem[Arora {\em et~al.\/}(2002)Arora, Sureshkumar \& Khomami]{arora2002}
{\sc \au{Arora, K}, \au{Sureshkumar, R} \& \au{Khomami, B}} \yr{2002}
  \at{Experimental investigation of purely elastic instabilities in periodic
  flows}.  \jt{Journal of non-newtonian fluid mechanics}  \bvol{108}~(1-3),
  \pg{209--226}.

\bibitem[Avgousti \& Beris(1993)]{avgousti1993}
{\sc \au{Avgousti, Marios} \& \au{Beris, Antony~N}} \yr{1993}
  \at{Non-axisymmetric modes in viscoelastic taylor-couette flow}.  \jt{Journal
  of non-newtonian fluid mechanics}  \bvol{50}~(2-3),  \pg{225--251}.

\bibitem[Babayekhorasani {\em et~al.\/}(2016)Babayekhorasani, Dunstan,
  Krishnamoorti \& Conrad]{babayekhorasani2016}
{\sc \au{Babayekhorasani, Firoozeh}, \au{Dunstan, Dave~E}, \au{Krishnamoorti,
  Ramanan} \& \au{Conrad, Jacinta~C}} \yr{2016}  \at{Nanoparticle dispersion in
  disordered porous media with and without polymer additives}.  \jt{Soft
  Matter}  \bvol{12}~(26),  \pg{5676--5683}.

\bibitem[Balkovsky {\em et~al.\/}(2000)Balkovsky, Fouxon \&
  Lebedev]{balkovsky2000}
{\sc \au{Balkovsky, E}, \au{Fouxon, A} \& \au{Lebedev, V}} \yr{2000}
  \at{Turbulent dynamics of polymer solutions}.  \jt{Physical review letters}
  \bvol{84}~(20),  \pg{4765}.

\bibitem[Batchelor(1971)]{batchelor1971}
{\sc \au{Batchelor, GK}} \yr{1971}  \at{The stress generated in a non-dilute
  suspension of elongated particles by pure straining motion}.  \jt{Journal of
  Fluid Mechanics}  \bvol{46}~(4),  \pg{813--829}.

\bibitem[Bernabe(1991)]{Bernabe1991}
{\sc \au{Bernabe, Y.}} \yr{1991}  \at{{Pore geometry and pressure dependence of
  the transport properties in sandstones}}.  \jt{Geophysics}  \bvol{56}~(4),
  \pg{424--576}.

\bibitem[Boger(1987)]{boger1987}
{\sc \au{Boger, DV}} \yr{1987}  \at{Viscoelastic flows through contractions}.
  \jt{Annual review of fluid mechanics}  \bvol{19}~(1),  \pg{157--182}.

\bibitem[Chertkov(2000)]{chertkov2000}
{\sc \au{Chertkov, Michael}} \yr{2000}  \at{Polymer stretching by turbulence}.
  \jt{Physical review letters}  \bvol{84}~(20),  \pg{4761}.

\bibitem[Clarke {\em et~al.\/}(2016)Clarke, Howe, Mitchell, Staniland, Hawkes
  {\em et~al.\/}]{clarke2016}
{\sc \au{Clarke, Andrew}, \au{Howe, Andrew~M}, \au{Mitchell, Jonathan},
  \au{Staniland, John}, \au{Hawkes, Laurence~A} \& \au{others}} \yr{2016}
  \at{How viscoelastic-polymer flooding enhances displacement efficiency}.
  \jt{SPE Journal}  \bvol{21}~(03),  \pg{675--687}.

\bibitem[Clasen {\em et~al.\/}(2006)Clasen, Plog, Kulicke, Owens, Macosko,
  Scriven, Verani \& McKinley]{clasen2006}
{\sc \au{Clasen, C.}, \au{Plog, J.~P.}, \au{Kulicke, W.-M.}, \au{Owens, M.},
  \au{Macosko, C.}, \au{Scriven, L.~E.}, \au{Verani, M.} \& \au{McKinley,
  G.~H.}} \yr{2006}  \at{{How dilute are dilute solutions in extensional
  flows?}}  \jt{Journal of Rheology}  \bvol{50},  \pg{849--881}.

\bibitem[De {\em et~al.\/}(2018{\natexlab{{\em a\/}}})De, Koesen, Maitri,
  Golombok, Padding \& van Santvoort]{de2018flow}
{\sc \au{De, S}, \au{Koesen, SP}, \au{Maitri, RV}, \au{Golombok, M},
  \au{Padding, JT} \& \au{van Santvoort, JFM}} \yr{2018{\natexlab{{\em a\/}}}}
  \at{Flow of viscoelastic surfactants through porous media}.  \jt{AIChE
  Journal}  \bvol{64}~(2),  \pg{773--781}.

\bibitem[De {\em et~al.\/}(2018{\natexlab{{\em b\/}}})De, Krishnan, van~der
  Schaaf, Kuipers, Peters \& Padding]{de2018viscoelastic}
{\sc \au{De, S}, \au{Krishnan, P}, \au{van~der Schaaf, J}, \au{Kuipers, JAM},
  \au{Peters, EAJF} \& \au{Padding, JT}} \yr{2018{\natexlab{{\em b\/}}}}
  \at{Viscoelastic effects on residual oil distribution in flows through
  pillared microchannels}.  \jt{Journal of colloid and interface science}
  \bvol{510},  \pg{262--271}.

\bibitem[De {\em et~al.\/}(2017{\natexlab{{\em a\/}}})De, Kuipers, Peters \&
  Padding]{de2017b}
{\sc \au{De, S}, \au{Kuipers, JAM}, \au{Peters, EAJF} \& \au{Padding, JT}}
  \yr{2017{\natexlab{{\em a\/}}}}  \at{Viscoelastic flow simulations in model
  porous media}.  \jt{Physical Review Fluids}  \bvol{2}~(5),  \pg{053303}.

\bibitem[De {\em et~al.\/}(2017{\natexlab{{\em b\/}}})De, Kuipers, Peters \&
  Padding]{de2017a}
{\sc \au{De, S}, \au{Kuipers, JAM}, \au{Peters, EAJF} \& \au{Padding, JT}}
  \yr{2017{\natexlab{{\em b\/}}}}  \at{Viscoelastic flow simulations in random
  porous media}.  \jt{Journal of Non-Newtonian Fluid Mechanics}  \bvol{248},
  \pg{50--61}.

\bibitem[De {\em et~al.\/}(2017{\natexlab{{\em c\/}}})De, Kuipers, Peters \&
  Padding]{de2017viscoelastic}
{\sc \au{De, Shauvik}, \au{Kuipers, Johannes~AM}, \au{Peters, Elias~AJF} \&
  \au{Padding, Johan~T}} \yr{2017{\natexlab{{\em c\/}}}}  \at{Viscoelastic flow
  past mono-and bidisperse random arrays of cylinders: flow resistance,
  topology and normal stress distribution}.  \jt{Soft matter}  \bvol{13}~(48),
  \pg{9138--9146}.

\bibitem[De {\em et~al.\/}(2016)De, van~der Schaaf, Deen, Kuipers, Peters \&
  Padding]{de2016}
{\sc \au{De, S}, \au{van~der Schaaf, J}, \au{Deen, NG}, \au{Kuipers, JAM},
  \au{Peters, EAJF} \& \au{Padding, JT}} \yr{2016}  \at{Elastic instabilities
  in flows through pillared micro channels}.  \jt{arXiv preprint
  arXiv:1607.03672} .

\bibitem[Doyen(1988)]{Doyen1988}
{\sc \au{Doyen, Philippe~M.}} \yr{1988}  \at{{Permeability, Conductivity, and
  Pore Geometry of Sandstone}}.  \jt{Journal of Geophysical Research}
  \bvol{93}~(B7),  \pg{7729--7740}.

\bibitem[Durst {\em et~al.\/}(1981)Durst, Haas \& Kaczmar]{durst1981}
{\sc \au{Durst, FRBU}, \au{Haas, R} \& \au{Kaczmar, BU}} \yr{1981}  \at{Flows
  of dilute hydrolyzed polyacrylamide solutions in porous media under various
  solvent conditions}.  \jt{Journal of Applied Polymer Science}  \bvol{26}~(9),
   \pg{3125--3149}.

\bibitem[Fran{\c{c}}ois {\em et~al.\/}(2009)Fran{\c{c}}ois, Amarouchene, Lounis
  \& Kellay]{francois2009}
{\sc \au{Fran{\c{c}}ois, Nicolas}, \au{Amarouchene, Yacine}, \au{Lounis,
  Brahim} \& \au{Kellay, Hamid}} \yr{2009}  \at{Polymer conformations and
  hysteretic stresses in nonstationary flows of polymer solutions}.  \jt{EPL
  (Europhysics Letters)}  \bvol{86}~(3),  \pg{34002}.

\bibitem[Galindo-Rosales {\em et~al.\/}(2012)Galindo-Rosales, Campo-Dea{\~n}o,
  Pinho, Van~Bokhorst, Hamersma, Oliveira \& Alves]{galindo2012}
{\sc \au{Galindo-Rosales, Francisco~J}, \au{Campo-Dea{\~n}o, Laura}, \au{Pinho,
  FT}, \au{Van~Bokhorst, E}, \au{Hamersma, PJ}, \au{Oliveira, M{\'o}nica~SN} \&
  \au{Alves, MA}} \yr{2012}  \at{Microfluidic systems for the analysis of
  viscoelastic fluid flow phenomena in porous media}.  \jt{Microfluidics and
  nanofluidics}  \bvol{12}~(1-4),  \pg{485--498}.

\bibitem[Galindo-Rosales {\em et~al.\/}(2014)Galindo-Rosales, Campo-Dea{\~n}o,
  Sousa, Ribeiro, Oliveira, Alves \& Pinho]{galindo2014}
{\sc \au{Galindo-Rosales, Francisco~J}, \au{Campo-Dea{\~n}o, Laura}, \au{Sousa,
  Patr{\'\i}cia~C}, \au{Ribeiro, Vera~M}, \au{Oliveira, M{\'o}nica~SN},
  \au{Alves, Manuel~A} \& \au{Pinho, Fernando~T}} \yr{2014}  \at{Viscoelastic
  instabilities in micro-scale flows}.  \jt{Experimental Thermal and Fluid
  Science}  \bvol{59},  \pg{128--139}.

\bibitem[Gerashchenko {\em et~al.\/}(2005)Gerashchenko, Chevallard \&
  Steinberg]{gerashchenko2005}
{\sc \au{Gerashchenko, S}, \au{Chevallard, C} \& \au{Steinberg, V}} \yr{2005}
  \at{Single-polymer dynamics: Coil-stretch transition in a random flow}.
  \jt{EPL (Europhysics Letters)}  \bvol{71}~(2),  \pg{221}.

\bibitem[Groisman \& Steinberg(2000)]{groisman2000}
{\sc \au{Groisman, Alexander} \& \au{Steinberg, Victor}} \yr{2000}  \at{Elastic
  turbulence in a polymer solution flow}.  \jt{Nature}  \bvol{405}~(6782),
  \pg{53}.

\bibitem[Gulati {\em et~al.\/}(2015)Gulati, Muller \& Liepmann]{gulati2015}
{\sc \au{Gulati, Shelly}, \au{Muller, Susan~J} \& \au{Liepmann, Dorian}}
  \yr{2015}  \at{Flow of dna solutions in a microfluidic gradual contraction}.
  \jt{Biomicrofluidics}  \bvol{9}~(5),  \pg{054102}.

\bibitem[Gupta {\em et~al.\/}(2004)Gupta, Sureshkumar \& Khomami]{gupta2004}
{\sc \au{Gupta, VK}, \au{Sureshkumar, R} \& \au{Khomami, B}} \yr{2004}
  \at{Polymer chain dynamics in newtonian and viscoelastic turbulent channel
  flows}.  \jt{Physics of Fluids}  \bvol{16}~(5),  \pg{1546--1566}.

\bibitem[Harnett \& Irvine(1979)]{harnett1979}
{\sc \au{Harnett, James~P} \& \au{Irvine, Thomas~F}} \yr{1979} {\em Advances in
  heat transfer\/}.  \publ{Elsevier Science}.

\bibitem[Haward {\em et~al.\/}(2016)Haward, McKinley \&
  Shen]{haward2016elastic}
{\sc \au{Haward, Simon~J}, \au{McKinley, Gareth~H} \& \au{Shen, Amy~Q}}
  \yr{2016}  \at{Elastic instabilities in planar elongational flow of
  monodisperse polymer solutions}.  \jt{Scientific reports}  \bvol{6},
  \pg{33029}.

\bibitem[Haward \& Odell(2003)]{haward2003}
{\sc \au{Haward, Simon~J} \& \au{Odell, Jeffrey~A}} \yr{2003}  \at{Viscosity
  enhancement in non-newtonian flow of dilute polymer solutions through
  crystallographic porous media}.  \jt{Rheologica acta}  \bvol{42}~(6),
  \pg{516--526}.

\bibitem[Haward {\em et~al.\/}(2018)Haward, Toda-Peters \& Shen]{haward2018}
{\sc \au{Haward, Simon~J}, \au{Toda-Peters, Kazumi} \& \au{Shen, Amy~Q}}
  \yr{2018}  \at{Steady viscoelastic flow around high-aspect-ratio,
  low-blockage-ratio microfluidic cylinders}.  \jt{Journal of Non-Newtonian
  Fluid Mechanics}  \bvol{254},  \pg{23--35}.

\bibitem[Howe {\em et~al.\/}(2015)Howe, Clarke \& Giernalczyk]{howe2015}
{\sc \au{Howe, Andrew~M}, \au{Clarke, Andrew} \& \au{Giernalczyk, Daniel}}
  \yr{2015}  \at{Flow of concentrated viscoelastic polymer solutions in porous
  media: effect of mw and concentration on elastic turbulence onset in various
  geometries}.  \jt{Soft Matter}  \bvol{11}~(32),  \pg{6419--6431}.

\bibitem[Huh {\em et~al.\/}(2008)Huh, Pope {\em et~al.\/}]{huh2008}
{\sc \au{Huh, Chun}, \au{Pope, Gary~Arnold} \& \au{others}} \yr{2008} Residual
  oil saturation from polymer floods: laboratory measurements and theoretical
  interpretation.  \bt{In {\em SPE Symposium on Improved Oil Recovery\/}}.
  Society of Petroleum Engineers.

\bibitem[Ioannidis \& Chatzis(1993)]{Ioannidis1993}
{\sc \au{Ioannidis, Marios~A.} \& \au{Chatzis, Ioannis}} \yr{1993}
  \at{{Network modelling of pore structure and transport properties of porous
  media}}.  \jt{Chemical Engineering Science}  \bvol{48}~(5),  \pg{951--972}.

\bibitem[Kawale {\em et~al.\/}(2017{\natexlab{{\em a\/}}})Kawale, Bouwman,
  Sachdev, Zitha, Kreutzer, Rossen \& Boukany]{kawale2017b}
{\sc \au{Kawale, Durgesh}, \au{Bouwman, Gelmer}, \au{Sachdev, Shaurya},
  \au{Zitha, Pacelli~LJ}, \au{Kreutzer, Michiel~T}, \au{Rossen, William~R} \&
  \au{Boukany, Pouyan~E}} \yr{2017{\natexlab{{\em a\/}}}}  \at{Polymer
  conformation during flow in porous media}.  \jt{Soft matter}  \bvol{13}~(46),
   \pg{8745--8755}.

\bibitem[Kawale {\em et~al.\/}(2017{\natexlab{{\em b\/}}})Kawale, Marques,
  Zitha, Kreutzer, Rossen \& Boukany]{kawale2017a}
{\sc \au{Kawale, Durgesh}, \au{Marques, Esteban}, \au{Zitha, Pacelli~LJ},
  \au{Kreutzer, Michiel~T}, \au{Rossen, William~R} \& \au{Boukany, Pouyan~E}}
  \yr{2017{\natexlab{{\em b\/}}}}  \at{Elastic instabilities during the flow of
  hydrolyzed polyacrylamide solution in porous media: Effect of pore-shape and
  salt}.  \jt{Soft matter}  \bvol{13}~(4),  \pg{765--775}.

\bibitem[Kenney {\em et~al.\/}(2013)Kenney, Poper, Chapagain \&
  Christopher]{kenney2013}
{\sc \au{Kenney, Stephen}, \au{Poper, Kade}, \au{Chapagain, Ganesh} \&
  \au{Christopher, Gordon~F}} \yr{2013}  \at{Large deborah number flows around
  confined microfluidic cylinders}.  \jt{Rheologica Acta}  \bvol{52}~(5),
  \pg{485--497}.

\bibitem[Khomami \& Moreno(1997)]{khomami1997}
{\sc \au{Khomami, Bamin} \& \au{Moreno, Luis~D}} \yr{1997}  \at{Stability of
  viscoelastic flow around periodic arrays of cylinders}.  \jt{Rheologica acta}
   \bvol{36}~(4),  \pg{367--383}.

\bibitem[Koelling \& Prud'homme(1991)]{koelling1991}
{\sc \au{Koelling, KW} \& \au{Prud'homme, Robert~Krafft}} \yr{1991}
  \at{Instabilities in multi-hole converging flow of viscoelastic fluids}.
  \jt{Rheologica acta}  \bvol{30}~(6),  \pg{511--522}.

\bibitem[Kwiecien {\em et~al.\/}(1990)Kwiecien, Macdonald \&
  Dullien]{Kwiecien1990}
{\sc \au{Kwiecien, M.~J.}, \au{Macdonald, I.~F.} \& \au{Dullien, F.~A.L.}}
  \yr{1990}  \at{{Three‐dimensional reconstruction of porous media from
  serial section data}}.  \jt{Journal of Microscopy}  \bvol{159}~(3),
  \pg{343--359}.

\bibitem[Lanzaro {\em et~al.\/}(2017)Lanzaro, Corbett \& Yuan]{lanzaro2017}
{\sc \au{Lanzaro, Alfredo}, \au{Corbett, Daniel} \& \au{Yuan, Xue-Feng}}
  \yr{2017}  \at{Non-linear dynamics of semi-dilute paam solutions in a
  microfluidic 3d cross-slot flow geometry}.  \jt{Journal of Non-Newtonian
  Fluid Mechanics}  \bvol{242},  \pg{57--65}.

\bibitem[Lanzaro {\em et~al.\/}(2015)Lanzaro, Li \& Yuan]{lanzaro2015}
{\sc \au{Lanzaro, Alfredo}, \au{Li, Zhuo} \& \au{Yuan, Xue-Feng}} \yr{2015}
  \at{Quantitative characterization of high molecular weight polymer solutions
  in microfluidic hyperbolic contraction flow}.  \jt{Microfluidics and
  Nanofluidics}  \bvol{18}~(5-6),  \pg{819--828}.

\bibitem[Lanzaro \& Yuan(2011)]{lanzaro2011}
{\sc \au{Lanzaro, Alfredo} \& \au{Yuan, Xue-Feng}} \yr{2011}  \at{Effects of
  contraction ratio on non-linear dynamics of semi-dilute, highly polydisperse
  paam solutions in microfluidics}.  \jt{Journal of Non-Newtonian Fluid
  Mechanics}  \bvol{166}~(17-18),  \pg{1064--1075}.

\bibitem[Larson(1992)]{larson1992}
{\sc \au{Larson, RG}} \yr{1992}  \at{Flow-induced mixing, demixing, and phase
  transitions in polymeric fluids}.  \jt{Rheologica Acta}  \bvol{31}~(6),
  \pg{497--520}.

\bibitem[McKinley {\em et~al.\/}(1996)McKinley, Pakdel \&
  {\"O}ztekin]{mckinley1996}
{\sc \au{McKinley, Gareth~H}, \au{Pakdel, Peyman} \& \au{{\"O}ztekin,
  Alparslan}} \yr{1996}  \at{Rheological and geometric scaling of purely
  elastic flow instabilities}.  \jt{Journal of Non-Newtonian Fluid Mechanics}
  \bvol{67},  \pg{19--47}.

\bibitem[Mongruel \& Cloitre(1995)]{mongruel1995}
{\sc \au{Mongruel, A} \& \au{Cloitre, M}} \yr{1995}  \at{Extensional flow of
  semidilute suspensions of rod-like particles through an orifice}.
  \jt{Physics Of Fluids}  \bvol{7}~(11),  \pg{2546--2552}.

\bibitem[Mongruel \& Cloitre(2003)]{mongruel2003}
{\sc \au{Mongruel, A} \& \au{Cloitre, M}} \yr{2003}  \at{Axisymmetric orifice
  flow for measuring the elongational viscosity of semi-rigid polymer
  solutions}.  \jt{Journal of non-newtonian fluid mechanics}  \bvol{110}~(1),
  \pg{27--43}.

\bibitem[O'Connell {\em et~al.\/}(2019)O'Connell, Lu, Browne \&
  Datta]{leapfrog}
{\sc \au{O'Connell, Margaret~G}, \au{Lu, Nancy~B}, \au{Browne, Christopher~A}
  \& \au{Datta, Sujit~S}} \yr{2019}  \at{Cooperative size sorting of deformable
  particles in porous media}.  \jt{Soft Matter}  \bvol{15},  \pg{3620}.

\bibitem[Odell \& Haward(2006)]{odell2006}
{\sc \au{Odell, JA} \& \au{Haward, SJ}} \yr{2006}  \at{Viscosity enhancement in
  non-newtonian flow of dilute aqueous polymer solutions through
  crystallographic and random porous media}.  \jt{Rheologica acta}
  \bvol{45}~(6),  \pg{853--863}.

\bibitem[Pakdel \& McKinley(1996)]{pakdel1996}
{\sc \au{Pakdel, Peyman} \& \au{McKinley, Gareth~H}} \yr{1996}  \at{Elastic
  instability and curved streamlines}.  \jt{Physical Review Letters}
  \bvol{77}~(12),  \pg{2459}.

\bibitem[Pan {\em et~al.\/}(2013)Pan, Morozov, Wagner \& Arratia]{pan2013}
{\sc \au{Pan, L}, \au{Morozov, A}, \au{Wagner, C} \& \au{Arratia, PE}}
  \yr{2013}  \at{Nonlinear elastic instability in channel flows at low reynolds
  numbers}.  \jt{Physical review letters}  \bvol{110}~(17),  \pg{174502}.

\bibitem[Pearson(1976)]{pearson1976}
{\sc \au{Pearson, JRA}} \yr{1976}  \at{Instability in non-newtonian flow}.
  \jt{Annual Review of Fluid Mechanics}  \bvol{8}~(1),  \pg{163--181}.

\bibitem[Pilitsis \& Beris(1989)]{pilitsis1989}
{\sc \au{Pilitsis, Stergios} \& \au{Beris, Antony~N}} \yr{1989}
  \at{Calculations of steady-state viscoelastic flow in an undulating tube}.
  \jt{Journal of Non-Newtonian Fluid Mechanics}  \bvol{31}~(3),  \pg{231--287}.

\bibitem[Pilitsis \& Beris(1991)]{pilitsis1991}
{\sc \au{Pilitsis, Stergios} \& \au{Beris, Antony~N}} \yr{1991}
  \at{Viscoelastic flow in an undulating tube. part ii. effects of high
  elasticity, large amplitude of undulation and inertia}.  \jt{Journal of
  non-newtonian fluid mechanics}  \bvol{39}~(3),  \pg{375--405}.

\bibitem[Pitts {\em et~al.\/}(1995)Pitts, Campbell, Surkalo, Wyatt {\em
  et~al.\/}]{pitts1995}
{\sc \au{Pitts, Malcolm~J}, \au{Campbell, Tom~A}, \au{Surkalo, Harry},
  \au{Wyatt, Kon} \& \au{others}} \yr{1995}  \at{Polymer flood of the rapdan
  pool, saskatchewan, canada}.  \jt{SPE Reservoir Engineering}  \bvol{10}~(03),
   \pg{183--186}.

\bibitem[Qin \& Arratia(2017)]{qin2017}
{\sc \au{Qin, Boyang} \& \au{Arratia, Paulo~E}} \yr{2017}  \at{Characterizing
  elastic turbulence in channel flows at low reynolds number}.  \jt{Physical
  Review Fluids}  \bvol{2}~(8),  \pg{083302}.

\bibitem[Qin {\em et~al.\/}(2019)Qin, Salipante, Hudson \& Arratia]{qin2019}
{\sc \au{Qin, Boyang}, \au{Salipante, Paul~F}, \au{Hudson, Steven~D} \&
  \au{Arratia, Paulo~E}} \yr{2019}  \at{Upstream vortex and elastic wave in the
  viscoelastic flow around a confined cylinder}.  \jt{Journal of Fluid
  Mechanics}  \bvol{864}.

\bibitem[Ribeiro {\em et~al.\/}(2014)Ribeiro, Coelho, Pinho \&
  Alves]{ribeiro2014}
{\sc \au{Ribeiro, VM}, \au{Coelho, PM}, \au{Pinho, FT} \& \au{Alves, MA}}
  \yr{2014}  \at{Viscoelastic fluid flow past a confined cylinder:
  Three-dimensional effects and stability}.  \jt{Chemical engineering science}
  \bvol{111},  \pg{364--380}.

\bibitem[Rodd {\em et~al.\/}(2007)Rodd, Cooper-White, Boger \&
  McKinley]{rodd2007}
{\sc \au{Rodd, LE}, \au{Cooper-White, JJ}, \au{Boger, DV} \& \au{McKinley, GH}}
  \yr{2007}  \at{Role of the elasticity number in the entry flow of dilute
  polymer solutions in micro-fabricated contraction geometries}.  \jt{Journal
  of Non-Newtonian Fluid Mechanics}  \bvol{143}~(2-3),  \pg{170--191}.

\bibitem[Roote(1998)]{roote1998}
{\sc \au{Roote, DS}} \yr{1998}  \at{Technology status report: in situ
  flushing}.  \jt{Ground Water Remediation Technology Analysis Center
  (available at http://www. gwrtac. org)} .

\bibitem[Sadanandan \& Sureshkumar(2004)]{sadanandan2004}
{\sc \au{Sadanandan, B} \& \au{Sureshkumar, Radhakrishna}} \yr{2004}
  \at{Global linear stability analysis of viscoelastic flow through a periodic
  channel}.  \jt{Journal of non-newtonian fluid mechanics}  \bvol{122}~(1-3),
  \pg{55--67}.

\bibitem[Sandiford {\em et~al.\/}(1964)]{sandiford1964}
{\sc \au{Sandiford, BB} \& \au{others}} \yr{1964}  \at{Laboratory and field
  studies of water floods using polymer solutions to increase oil recoveries}.
  \jt{Journal of Petroleum Technology}  \bvol{16}~(08),  \pg{917--922}.

\bibitem[Schroeder {\em et~al.\/}(2003)Schroeder, Babcock, Shaqfeh \&
  Chu]{schroeder2003}
{\sc \au{Schroeder, Charles~M}, \au{Babcock, Hazen~P}, \au{Shaqfeh, Eric~SG} \&
  \au{Chu, Steven}} \yr{2003}  \at{Observation of polymer conformation
  hysteresis in extensional flow}.  \jt{Science}  \bvol{301}~(5639),
  \pg{1515--1519}.

\bibitem[Shi \& Christopher(2016)]{shi2016}
{\sc \au{Shi, Xueda} \& \au{Christopher, Gordon~F}} \yr{2016}  \at{Growth of
  viscoelastic instabilities around linear cylinder arrays}.  \jt{Physics of
  Fluids}  \bvol{28}~(12),  \pg{124102}.

\bibitem[Shi {\em et~al.\/}(2015)Shi, Kenney, Chapagain \&
  Christopher]{shi2015}
{\sc \au{Shi, Xueda}, \au{Kenney, Stephen}, \au{Chapagain, Ganesh} \&
  \au{Christopher, Gordon~F}} \yr{2015}  \at{Mechanisms of onset for moderate
  mach number instabilities of viscoelastic flows around confined cylinders}.
  \jt{Rheologica Acta}  \bvol{54}~(9-10),  \pg{805--815}.

\bibitem[Son(2007)]{son2007}
{\sc \au{Son, Younggon}} \yr{2007}  \at{Determination of shear viscosity and
  shear rate from pressure drop and flow rate relationship in a rectangular
  channel}.  \jt{Polymer}  \bvol{48}~(2),  \pg{632--637}.

\bibitem[Sorbie(2013)]{sorbie2013}
{\sc \au{Sorbie, Kenneth~S}} \yr{2013} {\em Polymer-improved oil recovery\/}.
  \publ{Springer Science \& Business Media}.

\bibitem[Sousa {\em et~al.\/}(2015)Sousa, Pinho, Oliveira \& Alves]{sousa2015}
{\sc \au{Sousa, PC}, \au{Pinho, FT}, \au{Oliveira, MSN} \& \au{Alves, MA}}
  \yr{2015}  \at{Purely elastic flow instabilities in microscale cross-slot
  devices}.  \jt{Soft matter}  \bvol{11}~(45),  \pg{8856--8862}.

\bibitem[Sureshkumar {\em et~al.\/}(1994)Sureshkumar, Beris \&
  Avgousti]{sureshkumar1994}
{\sc \au{Sureshkumar, Radhakrishna}, \au{Beris, Antony~N} \& \au{Avgousti,
  Marios}} \yr{1994}  \at{Non-axisymmetric subcritical bifurcations in
  viscoelastic taylor--couette flow}.  \jt{Proceedings of the Royal Society of
  London. Series A: Mathematical and Physical Sciences}  \bvol{447}~(1929),
  \pg{135--153}.

\bibitem[Sureshkumar {\em et~al.\/}(1997)Sureshkumar, Beris \&
  Handler]{sureshkumar1997}
{\sc \au{Sureshkumar, R}, \au{Beris, Antony~N} \& \au{Handler, Robert~A}}
  \yr{1997}  \at{Direct numerical simulation of the turbulent channel flow of a
  polymer solution}.  \jt{Physics of Fluids}  \bvol{9}~(3),  \pg{743--755}.

\bibitem[Talwar \& Khomami(1995)]{talwar1995}
{\sc \au{Talwar, Kapil~K} \& \au{Khomami, Bamin}} \yr{1995}  \at{Flow of
  viscoelastic fluids past periodic square arrays of cylinders: inertial and
  shear thinning viscosity and elasticity effects}.  \jt{Journal of
  Non-Newtonian Fluid Mechanics}  \bvol{57}~(2-3),  \pg{177--202}.

\bibitem[Terrapon {\em et~al.\/}(2004)Terrapon, Dubief, Moin, Shaqfeh \&
  Lele]{terrapon2004}
{\sc \au{Terrapon, VE}, \au{Dubief, Yves}, \au{Moin, Parviz}, \au{Shaqfeh,
  Eric~SG} \& \au{Lele, Sanjiva~K}} \yr{2004}  \at{Simulated polymer stretch in
  a turbulent flow using brownian dynamics}.  \jt{Journal of Fluid Mechanics}
  \bvol{504},  \pg{61--71}.

\bibitem[Vanapalli {\em et~al.\/}(2006)Vanapalli, Ceccio \&
  Solomon]{vanapalli2006}
{\sc \au{Vanapalli, Siva~A}, \au{Ceccio, Steven~L} \& \au{Solomon, Michael~J}}
  \yr{2006}  \at{Universal scaling for polymer chain scission in turbulence}.
  \jt{Proceedings of the National Academy of Sciences}  \bvol{103}~(45),
  \pg{16660--16665}.

\bibitem[Varshney \& Steinberg(2017)]{varshney2017}
{\sc \au{Varshney, Atul} \& \au{Steinberg, Victor}} \yr{2017}  \at{Elastic wake
  instabilities in a creeping flow between two obstacles}.  \jt{Physical Review
  Fluids}  \bvol{2}~(5),  \pg{051301}.

\bibitem[V{\'a}zquez-Quesada \& Ellero(2012)]{vazquez2012}
{\sc \au{V{\'a}zquez-Quesada, A} \& \au{Ellero, M}} \yr{2012}  \at{Sph
  simulations of a viscoelastic flow around a periodic array of cylinders
  confined in a channel}.  \jt{Journal of Non-Newtonian Fluid Mechanics}
  \bvol{167},  \pg{1--8}.

\bibitem[Vermolen {\em et~al.\/}(2014)Vermolen, Van~Haasterecht, Masalmeh {\em
  et~al.\/}]{vermolen2014}
{\sc \au{Vermolen, ECM}, \au{Van~Haasterecht, MJT}, \au{Masalmeh, SK} \&
  \au{others}} \yr{2014} A systematic study of the polymer visco-elastic effect
  on residual oil saturation by core flooding.  \bt{In {\em SPE EOR Conference
  at Oil and Gas West Asia\/}}. Society of Petroleum Engineers.

\bibitem[Walkama {\em et~al.\/}(2019)Walkama, Waisbord \& Guasto]{walkama2019}
{\sc \au{Walkama, Derek~M}, \au{Waisbord, Nicolas} \& \au{Guasto, Jeffrey~S}}
  \yr{2019}  \at{Disorder suppresses chaos in viscoelastic flows}.  \jt{arXiv
  preprint arXiv:1906.11868} .

\bibitem[Wang {\em et~al.\/}(2011)Wang, Wang, Xia {\em et~al.\/}]{wang2011}
{\sc \au{Wang, Demin}, \au{Wang, Gang}, \au{Xia, Huifen} \& \au{others}}
  \yr{2011} Large scale high visco-elastic fluid flooding in the field achieves
  high recoveries.  \bt{In {\em SPE Enhanced Oil Recovery Conference\/}}.
  Society of Petroleum Engineers.

\bibitem[Wei {\em et~al.\/}(2014)Wei, Romero-Zer{\'o}n \& Rodrigue]{wei2014}
{\sc \au{Wei, Bing}, \au{Romero-Zer{\'o}n, Laura} \& \au{Rodrigue, Denis}}
  \yr{2014}  \at{Oil displacement mechanisms of viscoelastic polymers in
  enhanced oil recovery (eor): a review}.  \jt{Journal of Petroleum Exploration
  and Production Technology}  \bvol{4}~(2),  \pg{113--121}.

\bibitem[Zaitoun {\em et~al.\/}(1998)Zaitoun, Bertin, Lasseux {\em
  et~al.\/}]{zaitoun1998}
{\sc \au{Zaitoun, A}, \au{Bertin, H}, \au{Lasseux, D} \& \au{others}} \yr{1998}
  Two-phase flow property modifications by polymer adsorption.  \bt{In {\em
  SPE/DOE improved oil recovery symposium\/}}. Society of Petroleum Engineers.

\bibitem[Zaitoun {\em et~al.\/}(1988)Zaitoun, Kohler {\em
  et~al.\/}]{zaitoun1988}
{\sc \au{Zaitoun, A}, \au{Kohler, N} \& \au{others}} \yr{1988} Two-phase flow
  through porous media: effect of an adsorbed polymer layer.  \bt{In {\em SPE
  Annual Technical Conference and Exhibition\/}}. Society of Petroleum
  Engineers.

\bibitem[Zilz {\em et~al.\/}(2012)Zilz, Poole, Alves, Bartolo, Levach{\'e} \&
  Lindner]{zilz2012}
{\sc \au{Zilz, J}, \au{Poole, RJ}, \au{Alves, MA}, \au{Bartolo, D},
  \au{Levach{\'e}, B} \& \au{Lindner, A}} \yr{2012}  \at{Geometric scaling of a
  purely elastic flow instability in serpentine channels}.  \jt{Journal of
  Fluid Mechanics}  \bvol{712},  \pg{203--218}.

\end{thebibliography}
\end{document}